\pdfoutput=1
\documentclass[twocolumn,twoside,slac_two,floatfix,a4paper]{revtex4}
\usepackage{graphicx}
\usepackage{fancyhdr}
\begin{document}

%Title of paper
\title{A phenomenological study of photon production in low energy neutrino nucleon scattering}
\preprint{LA-UR-09-05513}
% Repeat the \author .. \affiliation  etc. as needed
%
% \affiliation command applies to all authors since the last
% \affiliation command. The \affiliation command should follow the
% other information

\author{James Jenkins}
\author{T. Goldman}
\affiliation{Theoretical Division,  
Los Alamos National Laboratory, Los Alamos, NM 87545}

\begin{abstract}
Low energy photon production is an important background to many current and future precision neutrino experiments.  We present a phenomenological study of $t$-channel radiative corrections to neutral current neutrino nucleus scattering.  After introducing the relevant processes and phenomenological coupling constants, we will explore the derived energy and angular distributions as well as total cross-section predictions along their estimated uncertainties.  This is supplemented throughout with comments on possible experimental signatures and implications.  We conclude with a general discussion of the analysis in the context of complimentary methodologies.
\end{abstract}

%\maketitle must follow title, authors, abstract
\maketitle

\thispagestyle{fancy}

% body of paper here - Use proper section commands
% References should be done using the \cite, \ref, and \label commands
% Put \label in argument of \section for cross-referencing
%\section{\label{}}

%%%%%%%%%%%%%%%%%%%%%%%%%%%%%%%%%%
\section{Introduction}

Recent neutrino scattering experiments report signals with accuracies below the 1\% level. Such 
unprecedented sensitivities demand corresponding efforts to 
determine backgrounds. Radiative corrections are clearly expected 
at this level.  A proper understanding of induced photon production is especially critical for those experiments searching for electron neutrino appearance with non-magnetized detectors~\cite{LSND,MB_NuRes,MB_NuBarRes} where it is difficult to distinguish gamma radiation from electrons.  This is the case for many precision short baseline oscillation experiments.  Standard radiative corrections from final state photon
bremsstrahlung~\cite{CoherentProdPhotonByNu} and resonant $\Delta/N^*$ production~\cite{PhoteElPiResProd} have already been examined in the literature and are included in experimental Monte Carlo simulations~\cite{GEANT4,nuance,GG}.  Next generation magnetized detectors will alleviate some the uncertainties induced by this background~\cite{MicroBooneProposal}.

In what follows, we present a novel Standard Model contribution to $t$-channel neutral current photon production in neutrino nucleon scattering first introduced by us in \cite{Jenkins:2009uq}.  In contrast to the well known $s$-channel effects described above, our selected class of processes are less obviously connected to the external line 
quanta.  We consider both neutrino and anti-neutrino processes and our 
results may be extended to other similar interactions both in 
neutral and charged current scattering. Although our primary focus is on modest 
energies, our results are relativistically covariant and thus may 
be applied to any energy.  Of course, at high energies Regge trajectory 
generalizations of the meson exchanges are necessary which will naturally lead to a quark picture of the interaction.  

This paper is organized as follows.  In section \ref{sec:process} we introduce the dominant scattering process diagrams and phenomenologically derived coupling constants.  This is followed by a derivation of the scattering cross-section.  We show our numerical differential and total cross-section results in section \ref{sec:pheno} where we also point out the importance of interference effects.  We conclude in section \ref{sec:conclusions} with a brief summary and general discussion of our methodology.

\section{Process}\label{sec:process}

\subsection{Diagram and Couplings}\label{subsec:DiaCoup}

\begin{figure}
\includegraphics[scale=0.38]{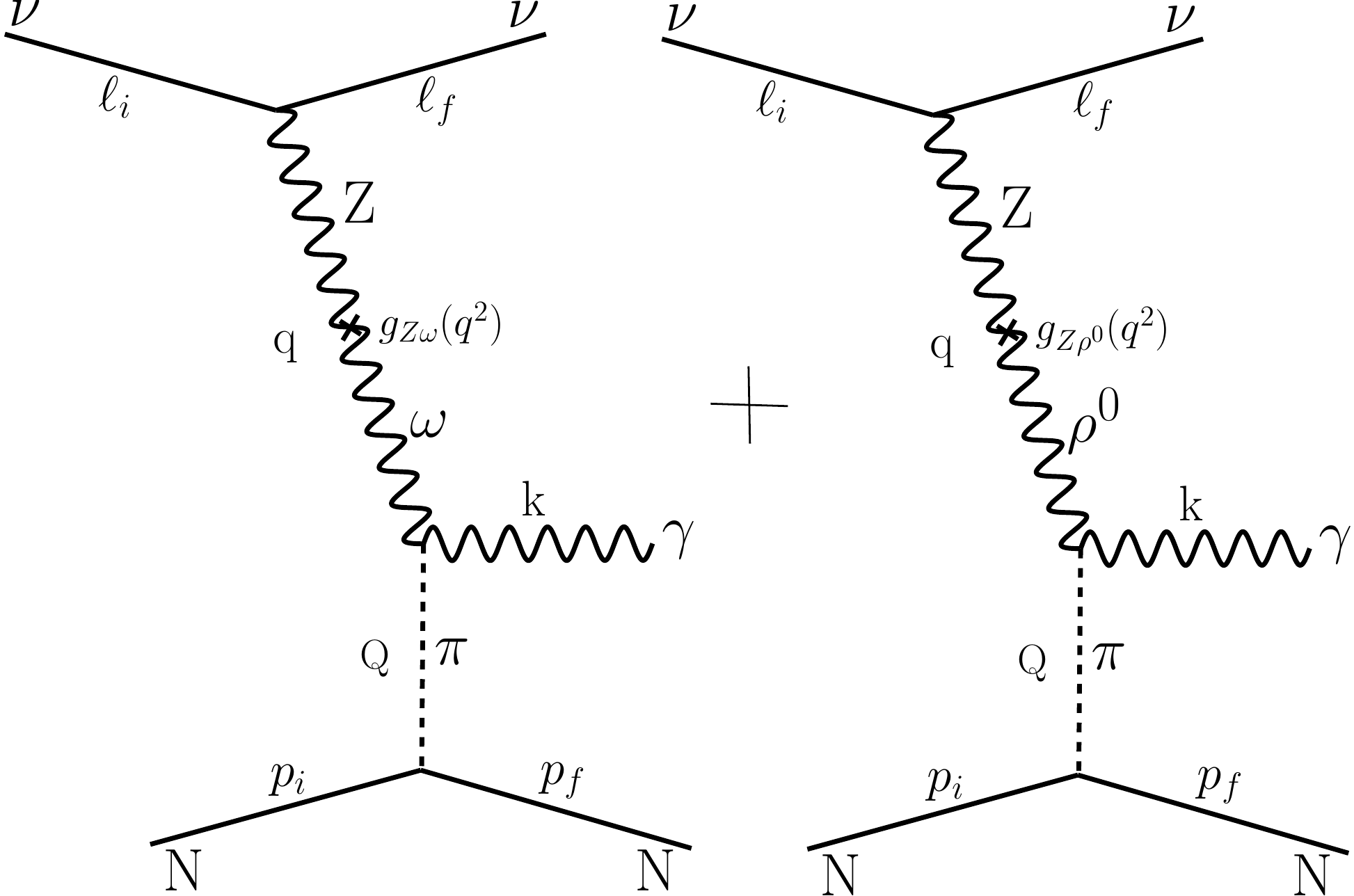}
\caption{Specific diagrams considered in this analysis.  Variants and interference effects are discussed 
in the text.}
\label{fig:NuNGammaDiag}
\end{figure}

Figure \ref{fig:NuNGammaDiag} shows the two dominant $t$-channel modes considered in this analysis, differentiated by intermediate $\omega$ and $\rho^0$ meson exchange.  In both diagrams, 
the Z-boson carrying the neutral current interaction from the neutrino line 
mixes into a vector boson in the familiar fashion of Vector Meson Dominance~\cite{vmd}.  The hadronic vector meson then undergoes a virtual decay to a photon and pion in the $t$-channel.  This last couples strongly to the hadron target.  Of course, there are other similar contributions from vector-meson (Regge) recurrences, but these predominantly affect only the 
overall strength for $q^2 \ll M^{\prime 2}$, where the excited state is integrated out of the interaction. Low energy hadron scattering experiments suggest that at modest 
energies the sum over all such contributions is likely to be dominated by these leading 
ones.  For the remainder of this section we will focus on the the $\omega$ exchange diagram and discuss the effects of interference in subsection \ref{subsec:interference}.

\begin{figure}
\includegraphics[scale=0.38]{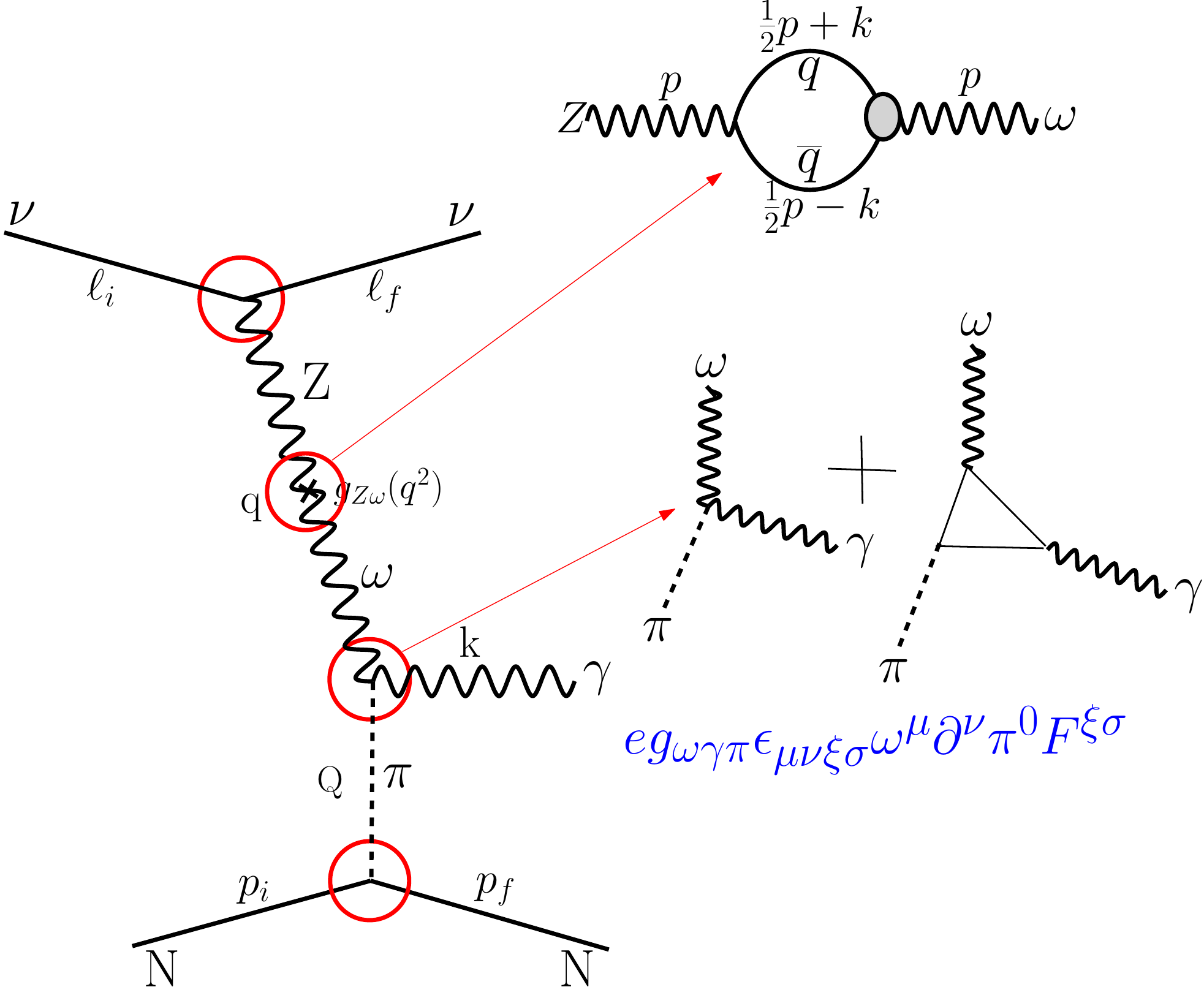}
\caption{Phenomenologically extracted coupling constants determined from experimental data.}
\label{fig:Couplings}
\end{figure}

The anatomy of this diagram is shown in figure \ref{fig:Couplings}, where the needed coupling constants are circled for emphasis.  These are extracted phenomenologically from measured processes. 

 Beginning with the $\pi-\gamma-$meson vertex.  We see that this contribution is similar to the triangle anomaly mediated interaction identified in \cite{H3} and discussed in \cite{Hill:2009ek}.  Our advantage over this approach is that the vertex strengths are known 
phenomenologically from the decay $\omega \rightarrow \pi^{0} + \gamma$ computed from the effective Lagrangian term 
\begin{equation}\label{eq:trilag}
{\cal L}_{I} = e g_{\omega\gamma\pi} \epsilon_{\mu\nu\xi \sigma} 
\omega^{\mu}\partial^{\nu}\pi^{0}F^{\xi \sigma}.  
\end{equation}
Here $F^{\xi \sigma}$ is the photon field strength tensor and the electromagnetic coupling $e$ is factored out for convenience since it is necessarily present from the photon interaction.  We point out that, although Eq.~(\ref{eq:trilag}) has the same form as that induced by the triangle anomaly due to the 
axial vector nature of the pion current, it exists independent of the anomaly.

Using this interaction, and neglecting the $\pi^0$ mass, we find the squared decay amplitude
\begin{equation}
  \mathcal{A}^2 = -\frac{2e^2g_{\omega\gamma\pi}^2}{3}k \cdot q = 
  \frac{e^2g_{\omega\gamma\pi}^2 M_\omega^4}{6}, 
\end{equation}
where $k$ and $q$ are the photon and pion momenta, respectively.  This implies the decay width
\begin{equation}\label{eq:decay}
 \Gamma(\omega \rightarrow \pi + \gamma) = \frac{\alpha g_{\omega\gamma\pi}^2 M_\omega^3}{24}.
\end{equation}
Fitting Eq.~(\ref{eq:decay}) to the observed decay width~\cite{PDG06}, we extract the coupling constant 
$g_{\omega\gamma\pi}=1.8/M_\omega$.  A similar exercise may be performed with 
the $\rho^0$ decay, in which case one extracts $g_{\rho\gamma\pi}=0.55/M_\rho$.  

The measured partial decay widths allow for very accurate coupling constant extraction beyond the $\mathcal{O}(10\%)$ level shown here.  For the purposes of describing a sub-$1\%$ signal, our accuracy adequately provides total cross-section predictions to better than $0.1\%$.  This reasoning holds for other parameter extractions given throughout the text.

  Moving on, the strength of the $Z-\omega$ 
mixing and its $p^2$ running may be extracted from the self energy diagram shown in figure \ref{fig:Couplings}.  Following \cite{GHT1,GHT2} we parameterize the $\omega-q-\bar{q}$ form factor by $g_{\omega q \bar{q}}M^2/(M^2-p^2)$ where $M$ describes the finite size of the $\omega$ meson.  The bare $\omega-q-\bar{q}$ coupling is found to be $g_{\omega q \bar{q}}~\approx~3.1$ from the decay $\omega~\rightarrow~\pi^0~\pi^+~\pi^-$ \cite{PDG06}.  Calculating the self energy via dimensional regularization and  considering only those terms that contribute to the $p^2$ dependence of $\omega-Z$ mixing we find 
\begin{eqnarray}\label{eq:wZrunning}
 &&g_{\omega Z}(p^2) = \frac{-g g_{\omega q\bar{q}}M^2 s^2_{W}}{12\pi^2
 c_{W}} \\ \nonumber
&\times& \int_0^1 dz \int_0^{1-z} dx \frac{p^2z(z-1) + m_q^2}{p^2z(z-1) + 
 m_q^2+x(M^2-m_q^2)},
\end{eqnarray}
after dropping logarithmically  
divergent contributions.  Here $c_{W}$ and $s_{W}$ are the cosine and sine of the weak mixing angle.  Taking reasonable limits of Eq.~(\ref{eq:wZrunning}) yields simplified analytic results \cite{Jenkins:2009uq} but the remaining Feynman integrals may be easily performed numerically.  Doing this, we find the averaged $\bar{g}_{\omega Z}~=~600~{\rm MeV}^2$ assuming $m_q \sim 3 \rm{MeV}$ and $M \sim M_\omega$ at momenta transfers between $200-1000 ~\rm{MeV}$.  We find a slight $\mathcal{O}(10\%)$ variation of $g_{\omega Z}(p^2)$ within this region of interest.

For the remaining couplings, we make use of the Standard Model weak interaction 
of the $Z$-boson to neutrinos and quarks and the well known pion coupling to the nucleon \cite{LatticePiNNFormFactor} via the interaction ${\cal L}_{I} = g_{\pi NN} \bar{\Psi}\gamma^{\mu}\gamma^5\partial_{\mu}\vec{\pi}\cdot \vec{\tau}\Psi$, where $\Psi$ is the nucleon field and $\vec{\tau}$ are isospin generators. No other parameters are required, so the prediction of the contribution to 
the total cross-section for producing a final state photon is absolute for 
this diagram.  The analogous analysis is easy to perform for the $\rho^0$ exchange case.

\subsection{Cross Section}

Evaluating the $\omega$ exchange diagram of figure \ref{fig:NuNGammaDiag}, we find the squared scattering amplitude
\begin{eqnarray}\label{eq:amplitude}
 \mathcal{A}^2 &=& \frac{128 M_N^2 g_{\nu Z}^2 g_{\omega\gamma\pi}^2 g_{\pi N N}^2 
 g_{\omega Z}^2(q^2)}{(q^2-M_Z^2)^2(q^2 - M_\omega^2)^2(Q^2-M_\pi^2)^2} \\
&\times& \ell_i 
 \cdot \ell_f (p_i \cdot p_f-M_N^2)\left((k \cdot \ell_i)^2 + (k \cdot \ell_f)^2 \right) \nonumber
\end{eqnarray}
in terms of the labeled four momenta.  The upper portion of Eq.~(\ref{eq:amplitude}) shows the general coupling constant and propagator dependencies while the lower factor describes the kinematics that follow from the diagram's Lorentz structure. In the Center of Mass (CM) frame, the momenta can be written explicitly as
\begin{eqnarray}
 \ell_i &=& (E_{\ell_i},\vec{E}_{\ell_i}) \\
p_i &=& (E_{p_i},-\vec{E}_{\ell_i}) \\
 \ell_f &=& (E_{\ell_f},\vec{E}_{\ell_f}) \\
p_f &=& (E_{p_f},\vec{p}_{p_f}) \\
 k &=& (E_{k},\vec{E}_{k}),
\end{eqnarray}
where we employ the shorthand $\vec{E}_{i} = \vec{p}_{i}$ to indicate a massless particle's 3-momentum of magnitude $E_i$.  In this frame we find, after performing trivial integrations over momentum-conserving delta functions, the differential cross-section for the final state photon's energy and angular distribution to be
\begin{eqnarray}\label{eq:diffsig1}
 &&\frac{d\sigma}{dE_kd\mu} = \frac{M_N^2g_{\nu Z}^2 g_{\omega\gamma\pi}^2 g_{\pi N N}^2}
 {(2\pi)^4 E_{\ell_i}(E_{\ell_i} + E_{p_i})} \\ \nonumber
&\times& \int dE_{\ell_f} d \phi \frac{g_{\omega Z}^2(q^2) q^2 
 Q^2 \left((k \cdot \ell_i)^2 + (k \cdot \ell_f)^2 \right)}{(q^2-M_Z^2)^2(q^2-M_\omega^2)^2 
 (Q^2 - M_\pi)^2},
\end{eqnarray}
where the momenta transfers are given by $q^2 = -2E_{\ell_i}E_{\ell_f}(1-\mu_{\ell_f})$ and $Q^2 = q^2 - 2k \cdot \ell_i + 2k \cdot \ell_f$.  Here $\mu = \cos\theta$ is the photon opening angle from the beam direction and $\mu_{\ell_f}$ is the cosine of the opening 
angle between the neutrino in the final and initial state.  It is related to $\mu$ and the cosine 
 of the opening angle between the photon and the final state neutrino $\mu_{\ell_f k}$ by  
\begin{equation}
\mu_{\ell_f} = \mu\mu_{\ell_f k} + \sqrt{1-\mu^2}\sqrt{1 - \mu_{\ell_f k}^2}\cos\phi.
\end{equation}
This is the only $\phi$ dependent term in the system.
Momentum conservation then fixes the remaining opening angle to be
\begin{eqnarray}
\mu_{\ell_f k} &=& \frac{1}{2E_{\ell_f}E_k} \\ \nonumber 
&\times&\left( s - 2\sqrt{s}(E_{\ell_f} + E_{k}) + 
2E_kE_{\ell_f} - M_N^2 \right),
\end{eqnarray}
where $s = (E_{\ell_i} + E_{p_i})^2$ is the relativistically invariant squared CM energy.
Additional constraints and limits of integration are found by requiring that $\mu_{\ell_f k}$ and $E_k$ take on physical values.

Evaluating Eq.~(\ref{eq:diffsig1}) subject to these constraints in the reasonable limit $|q^2| \ll M_Z^2$, $|Q^2| \gg M_\pi^2$ and $g_{\omega Z}(q^2) \sim \bar{g}_{\omega Z}$, we integrate over $\phi$ to obtain
\begin{eqnarray}\label{eq:diffsigma}
 &&\frac{d\sigma}{dE_kd\mu} = \frac{M_N^2 E_k^2 g_{\nu Z}^2\bar{g}_{\omega Z}^2 
 g_{\omega\gamma\pi}^2 g_{\pi N N}^2}{(2\pi)^3 E_{\ell_i}(E_{\ell_i} + E_{p_i}) 
 M_Z^4} \\ \nonumber
&\times&\int dE_{\ell_f} \left( E_{\ell_i}^2(1-\mu)^2 + E_{\ell_f}^2(1-\mu_{\ell_fk})^2 
 \right) \\ \nonumber
 &\times& \frac{1}{f^2(b-c)^2} \left\{ \frac{a-b}{(b^2-1)^{\frac{1}{2}}} + \frac{c^3 - 2ac^2 
 + abc - b + a}{(c^2 - 1)^{\frac{3}{2}}} \right\},
\end{eqnarray}
where
\begin{eqnarray}
 f &=& 2E_{\ell_i}E_{\ell_f}\sqrt{1-\mu^2}\sqrt{1-\mu_{\ell_fk}^2}\\
 a &=& \frac{2E_{\ell_i}E_{\ell_f}(1-\mu\mu_{\ell_fk})}{f}\\
 b &=& a + \frac{ 2E_kE_{\ell_i}(1-\mu) - 2E_k
 		E_{\ell_f}(1-\mu_{\ell_fk})}{f}\\
 c &=& a + \frac{M_\omega^2}{f}.
\end{eqnarray}
Assuming physical parameters, the dimensionless quantities $a,~b$ and $c$ are all greater than 
unity.  This leaves only the one-dimensional integral over the final neutrino energy ($E_{\ell_f}$) to perform. 

\section{Phenomenology}\label{sec:pheno}

In what follows, we numerically explore Eq.~(\ref{eq:diffsig1}) in both the CM and lab frames using the phenomenologically derived coupling constants.  We point out that we are using the full cross-section expression without approximation, including the $q^2$ running of $g_{\omega Z}(q^2)$.   We first discuss the results of the $\omega$ exchange process alone followed by the modifications induced by  
interference with
the contribution of the $\rho^{0}$ graph.. 

\subsection{Results}

\begin{figure}
\includegraphics[scale=0.17]{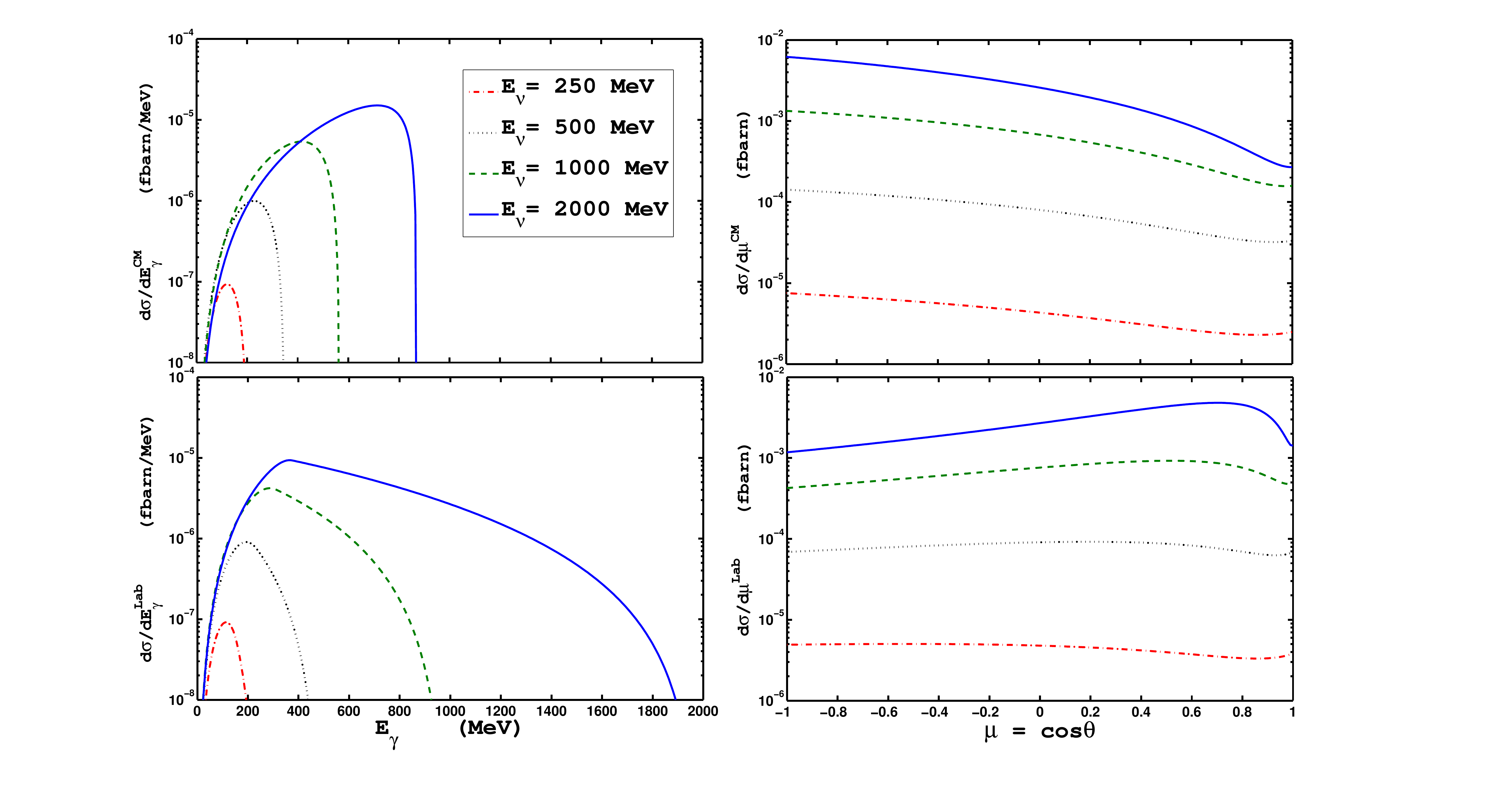}
\caption{Angular and energy differential cross-section distributions in the CM and lab frames for various neutrino beam energies.}\label{fig:Multisig}
\end{figure}

\begin{figure}
\includegraphics[scale=0.38]{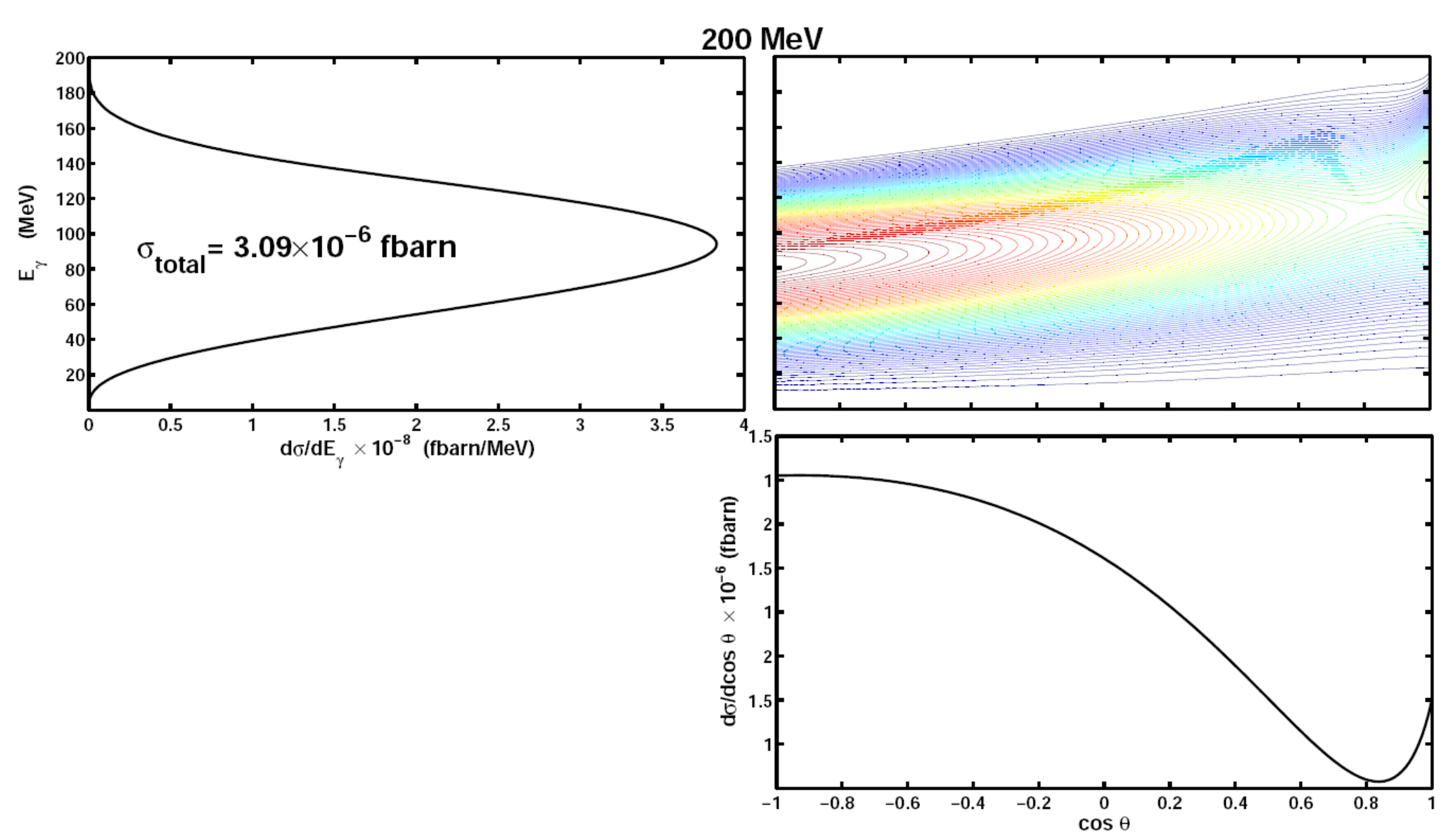}
\caption{Lab frame differential cross-section contour plot with beam energy $E_{\ell_i} = 200~{\rm MeV}$.  Angular and energetic projections are shown for convenience.  }
\label{fig:CS200MeV}
\end{figure}

\begin{figure}
\includegraphics[scale=0.38]{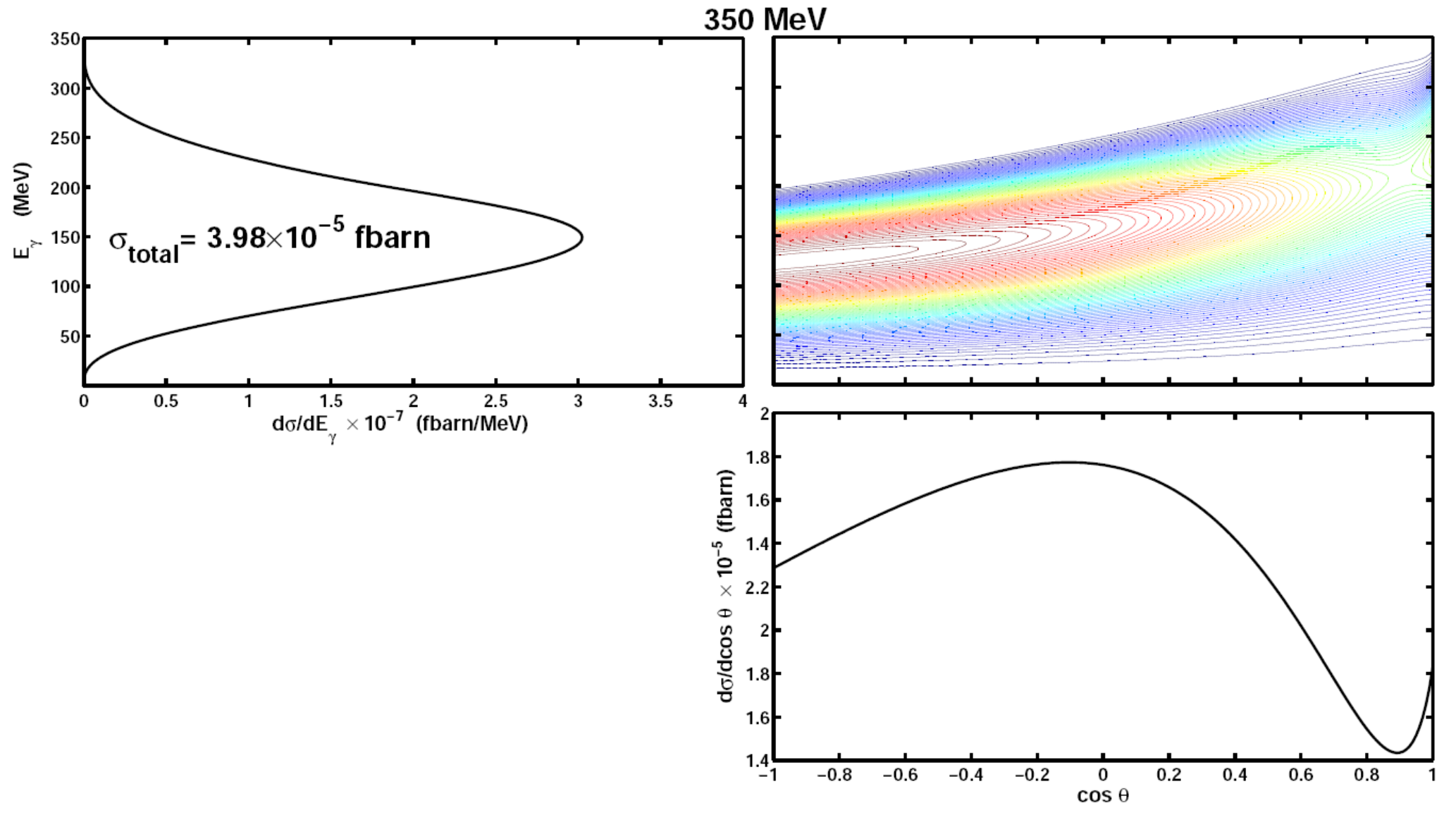}
\caption{Lab frame differential cross-section contour plot with beam energy $E_{\ell_i} = 350~{\rm MeV}$.  Angular and energetic projections are shown for convenience.  }
\label{fig:CS350MeV}
\end{figure}

\begin{figure}
\includegraphics[scale=0.38]{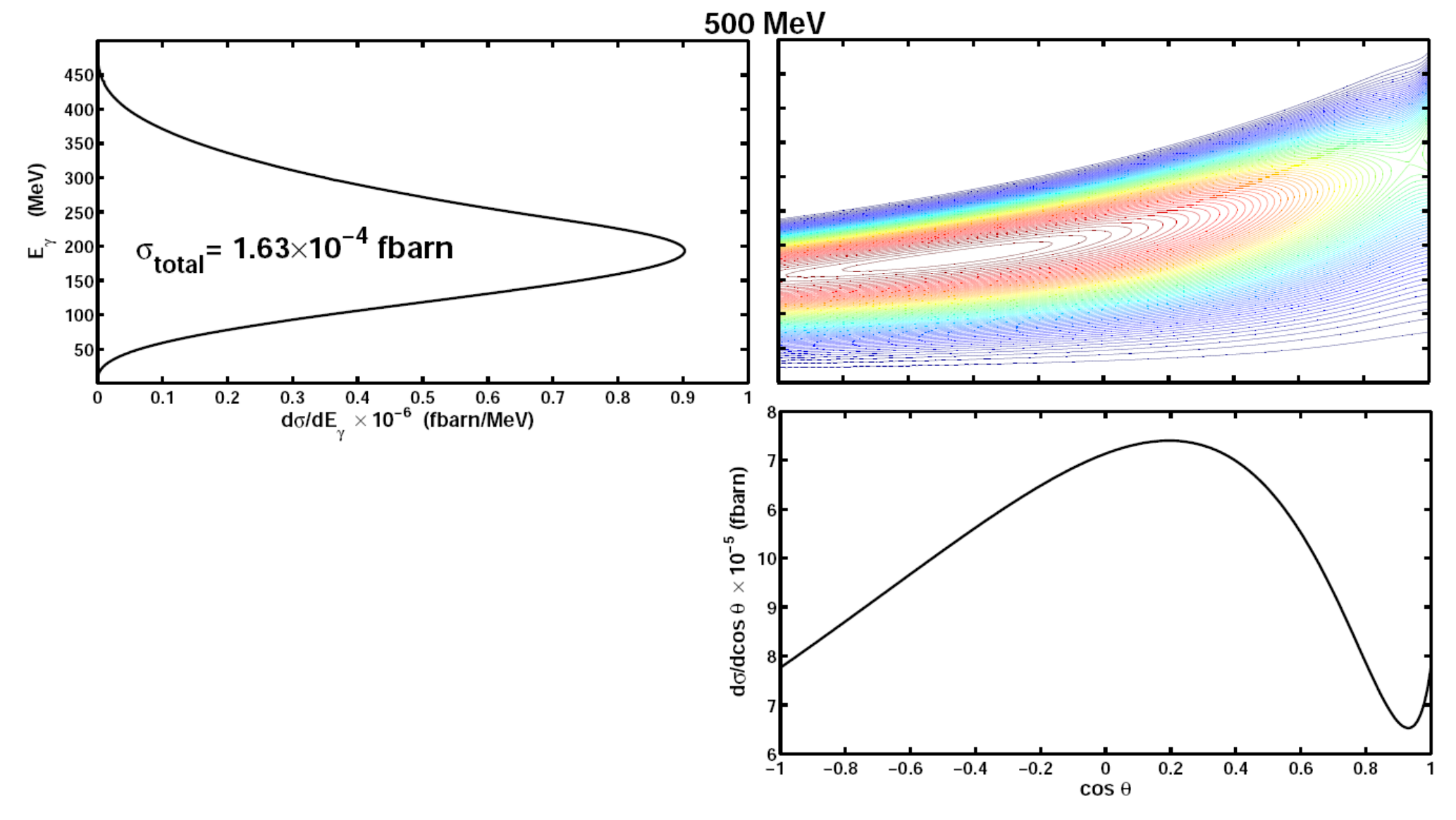}
\caption{Lab frame differential cross-section contour plot with beam energy $E_{\ell_i} = 500~{\rm MeV}$.  Angular and energetic projections are shown for convenience.  }
\label{fig:CS500MeV}
\end{figure}

\begin{figure}
\includegraphics[scale=0.38]{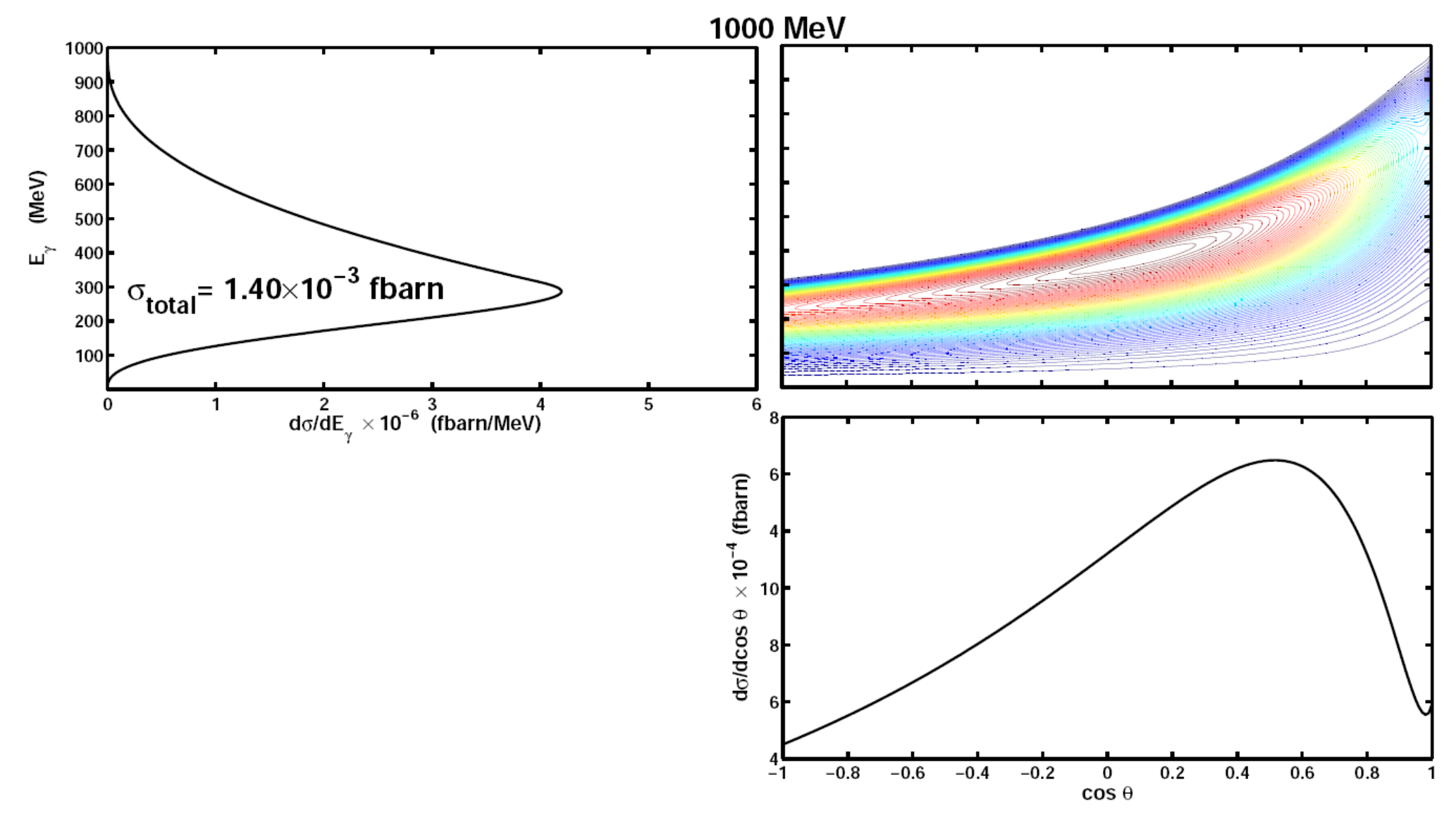}
\caption{Lab frame differential cross-section contour plot with beam energy $E_{\ell_i} = 1000~{\rm MeV}$.  Angular and energetic projections are shown for convenience.  }
\label{fig:CS1000MeV}
\end{figure}

In the CM frame, the predicted cross-section is weakly peaked in the backward direction 
with an energy maximum near the highest kinematically allowed energies due to the overall factor of 
$E_k$ in Eq.~(\ref{eq:diffsigma}).  This can be seen in the upper panels of figure \ref{fig:Multisig}.  Boosting these distributions to the lab frame pushes the angular distribution forward and spreads 
the energy of the photon, as is evident in the lower panels. Numerically integrating Eq.~(\ref{eq:diffsigma}), 
we plot the lab frame differential cross-section for beam energies of $200~{\rm MeV}$, 
$350~{\rm MeV}$, $500~{\rm MeV}$ and $1000~{\rm MeV}$ in figures \ref{fig:CS200MeV}, 
\ref{fig:CS350MeV}, \ref{fig:CS500MeV} and \ref{fig:CS1000MeV}, respectively.  In each case, we display the 
$E_\gamma$ and $\cos\theta$ dependent contour plots as well as energy and angular projection 
panels obtained by integrating over one of the variables.  The total cross-section 
is also noted for reference.  The angular distribution moves toward the forward peak 
with increasing neutrino energy due to the growing boosts from the CM to the lab frame.  
The distribution consistently peaks near the center of the kinematically allowed 
photon energy range.

\begin{figure}
\includegraphics[scale=0.17]{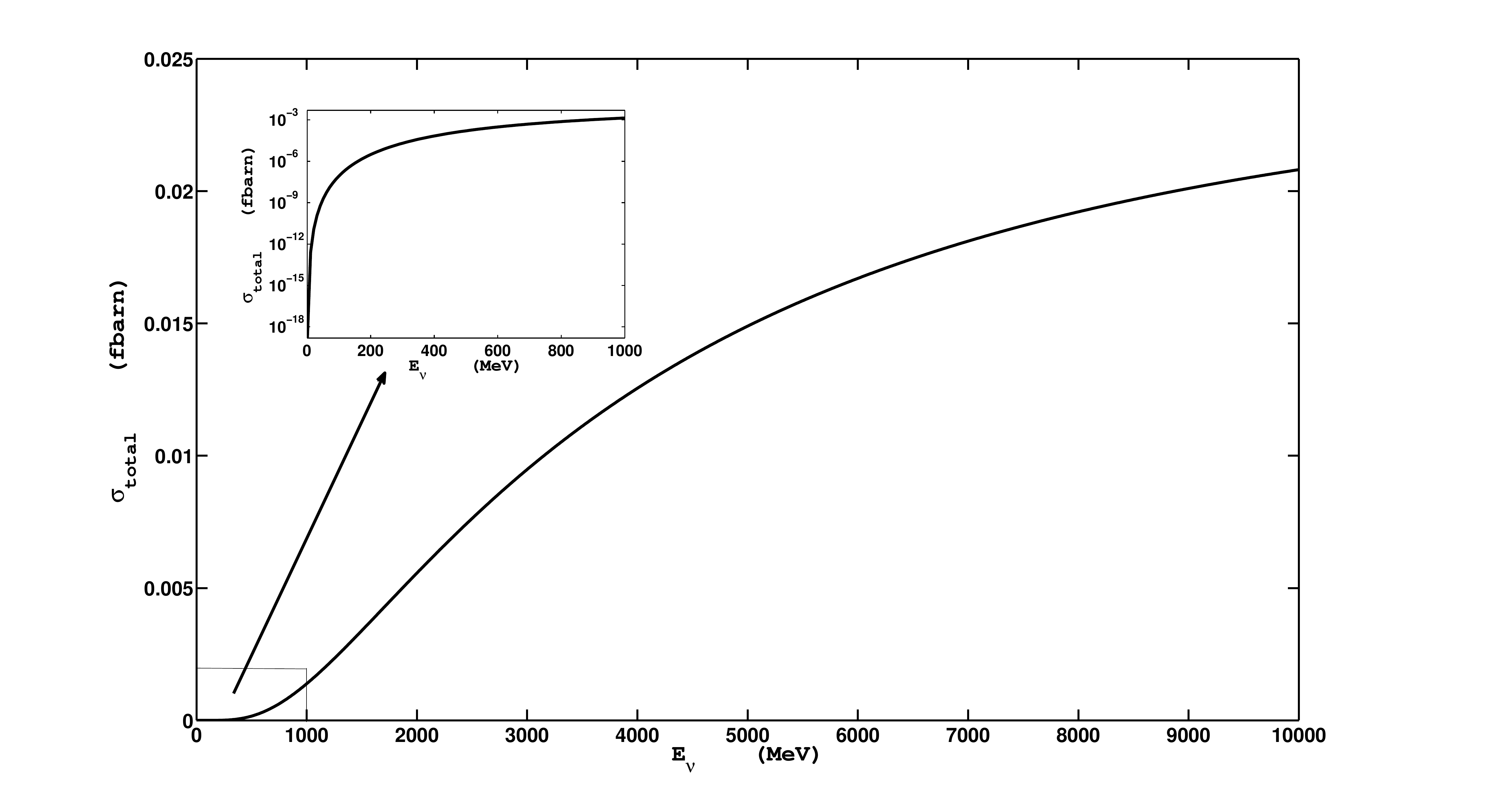}
\caption{Total cross-section as a function of neutrino beam energy.  A log scale insert plot is given to emphasize the low energy region of interest.}
\label{fig:TotalSig}
\end{figure}

Integrating over the final state photon energy and angular distribution, we plot the total 
cross-section as a function of neutrino beam energy in figure \ref{fig:TotalSig}.  At 
high energies the cross-section grows as $\sqrt{E_\nu}$ and near threshold as $E_\nu^2$.  
A logarithmic insert plot showing the low energy region of interest is included for convenience. Here, the cross-section is roughly three orders of magnitude smaller than the typical 
charged current cross-sections~\cite{Lipari:1994pz}.  However, this may still affect current~\cite{MB_NuRes,MB_NuBarRes} and future~\cite{MicroBooneProposal} experiments.  Additionally, long baseline and precision scattering neutrino experiments performed at higher energies 
(see, for example~\cite{MyNusongPRD,MINOS_Res08,NOvAProposal,T2KProspects,BNL_NuWorkGroup} and 
references therein) will be sensitive to this class of processes with an order of magnitude 
enhanced cross-section.

\subsection{Interference Effects}\label{subsec:interference}

We now discuss the interference effects resulting from the addition of the $\rho^0$ exchange diagram of figure \ref{fig:NuNGammaDiag}.  We point out that the $\rho^0$ exchange mode's scattering amplitude will have the same form as in the $\omega$ case with different (but still phenomenologically extracted) coupling constants and exchanged masses $M_\omega \leftrightarrow M_{\rho^0}$.  The couplings only effect the overall magnitude while the meson masses can also influence the cross-section shape.  Since $M_\omega \sim M_{\rho^0}$ \cite{PDG06} we see that the resulting spectral distributions should be almost identical.  Thus, it is sufficient to consider variations between the overall scattering magnitudes induced by coupling constant differences.

Such relative differences occur due to the $\pi-\gamma-$meson vertex as well as in the meson-$Z$ mixing term.  In the first case, the relevant coupling constants were calculated in subsection \ref{subsec:DiaCoup} leading to a suppression   
\begin{equation}
 \frac{g_{\rho\gamma\pi}}{g_{\omega\gamma\pi}} = \frac{0.55 M_\omega}{1.8 M_\rho} = 0.31.
\end{equation}

The meson-$Z$ mixing contribution is less trivial to understand.  On the basis of $SU(3)$ flavor symmetry, one expects similar results for the $\rho^0-Z$ and $\omega-Z$ mixing terms up to isospin effects. Looking at the self energy diagram in figure \ref{fig:Couplings} we see that the $\omega$ couples equally to the u and d quarks that contribute to the loop, whereas the $\rho^{0}$ does so with opposite signs due to isospin.  The standard model couplings are 
\begin{eqnarray}
g_{Zu\bar{u}} &=& \frac{g}{4c_{W}} (\frac{8}{3}s^2_{W}-1)  \\
g_{Zd\bar{d}} &=& \frac{g}{4c_{W}} (1-\frac{4}{3}s^2_{W}). 
\end{eqnarray}
 As might be expected from the fact that the $Z$-boson is dominantly isospin one
like the $\rho$, the $Z-\rho$ mixing is enhanced relative to the $Z-\omega$ mixing.  Combining this reasoning with $SU(3)$ flavor symmetry breaking manifest in deviations of $g_{\rho q\bar{q}}/g_{\omega q\bar{q}}$ from unity, we find
\begin{eqnarray}
&&\frac{g_{\rho Z}}{g_{\omega Z}} = \frac{g_{Z d\bar{d}} - g_{Z u\bar{u}}}{g_{Z d\bar{d}} +
 g_{Z u\bar{u}}} \times \frac{g_{\rho q\bar{q}}}{g_{\omega q\bar{q}}} \\
 &\sim&  3\frac{1-2s^2_{W}}{2s^2_{W}}\times \sqrt{\frac{\Gamma(\rho\rightarrow\pi\pi)}
{ \Gamma(\omega\rightarrow\pi\pi\pi)}  
 \frac{ \phi(\omega\rightarrow\pi\pi\pi)}{\phi(\rho\rightarrow\pi\pi)}} = 4.1, \nonumber
 \end{eqnarray}
where the $\phi$'s denote phase space integrals required for the meson$-q-\bar{q}$ coupling constant extractions.  

Thus, we estimate comparable cross-sections given by 
\begin{equation}
 \sigma^\rho/\sigma^\omega \approx 
(g_{\rho-\gamma-\pi}/g_{\omega-\gamma-\pi})^2 \times (g_{\rho Z}/g_{\omega Z})^2 \approx 
1.6, 
\end{equation}
which arises from the accidental cancellation of the meson$-\pi-\gamma$ suppression and the meson-$Z$ mixing enhancement.

The amplitudes for these processes are similar and significant interference is expected 
to occur.  From this effect the overall cross-sections may be modified by a factor between 
$0.07$ and $5.1$ for total destructive and constructive interference respectively.  Within the framework of the triangle anomaly \cite{H3} one may gain a handle on the relative interference phase by considering the low energy limit where the $\omega$ and $\rho$ contributions must sum to yield the $\pi^0-\gamma-Z$ coupling equal to $1-4s^2_{W}$.  This is small and picks out the lower bound of our interference region.  However, since the  
phase relations of
our phenomenological amplitudes are not fixed and are independent of  
the triangle
anomaly, isospin constraints from the quark couplings  
to the $Z$-boson
cannot be applied. This allows for additional latitude in matching  
experimental results.  
Cross-sections yielded by the lower limit fall well below expected future experimental 
sensitivities and therefore predict a  
negligible background. Contributions at the upper limit would have a substantial observable impact on precision measurements, and as such, should be included in future experimental Monte Carlos.

\section{Conclusions and Outlook}\label{sec:conclusions}

Other processes related to those in figure \ref{fig:NuNGammaDiag}, such as by the exchange 
of the $\omega$ and $\pi^0$ for other mesons (with the same quantum numbers), will 
contribute to similar production of photons in the final state.  These will differ from our 
calculation only by the coupling constants and meson masses which are necessarily heavier and should lead to propagator suppressions.  

After studying the diagrams in figure \ref{fig:NuNGammaDiag}, it is clear that this process class yields identical results for both neutrinos and anti-neutrinos.  The only difference between these amplitudes resides in a sign  
change at the axial-vector neutrino-Z coupling. This vanishes when Lorentz contracted with the rest of the 
diagram which is symmetric under the free indices at the vertex.  Additionally, we find 
by means of direct computation that many variants of figure \ref{fig:NuNGammaDiag} vanish due 
to similar symmetry reasons. In particular, amplitudes from diagrams with axial vector or pseudoscalar, as opposed to vector, meson exchange vanish.  Additionally, the ``reversed'' diagram, where the $\pi^0$ couples to the neutrino line, yields a null contribution.  We point out that such a contribution, if nonzero in principle, would be highly suppressed by the $\pi-\nu-\nu$ coupling.

Throughout this analysis we have used an on-shell coupling strength for the $\omega-
\pi-\gamma$ vertex, which is a commonly used phenomenological technique. 
The slow variation of the computed $Z-\omega$ mixing supports such 
an approach, but a three body interaction could behave differently. In this case one would expect a decrease in amplitude as vertex form factors act 
to suppress the effective coupling~\cite{offsh}. The vertex structure for the 
$\omega-q-\bar{q}$ coupling, also obtained from an on-shell decay, produces the same kind of uncertainties.  We believe that this issue is not a serious problem for our analysis, as the $Z-\omega$ mixing is only used within a few mass squared units from the on-shell point. 

Another potential background in $\nu_e$ appearance searches occurs when the a decay photon from a produced $\pi^0$ is lost to 
the detector, leaving a single photon faking an electron track. Fortunately, in this case the event rate can be normalized to the 
corresponding process in charged current neutrino scattering, which produces a 
neutral pion in conjunction with the charged lepton.  Although this may dominantly 
occur due to intermediate state processes, such as production of a $\Delta$ baryon 
or N$^{*}$ followed by its decay back to a nucleon and a pion (see 
\cite{SinglePiProdNuReact} and references therein), concern also arises regarding other 
processes, including those that may be coherent over the entire nuclear target with 
an amplified rate~\cite{EvidenceNuCoherentPiProd,CoherentPiProdNuReactions}. 
The preceding analysis may be applied, in an analogous way, to coherent $t$-channel pion production.  This 
parallel process is interesting from the point of view of interference. 
If the interference is destructive for photon production, there can be a significant difference between coherent pion 
production between charged and neutral current modes, as this interference cannot occur in 
the charged current case.  We will explore this possibility in a future study.

% If you have acknowledgments, this puts in the proper section head.
\bigskip % extra skip inserted
%%%%%%%%%%%%%%%%%%%%%%%%%%%%%%%%%%
\begin{acknowledgments}
J.J. thanks the organizers of DPF-2009 for the invitation to present this work as well as Richard Hill for useful discussions on the magnitude of the $\rho-\omega$ interference and insight into his approach to this problem. This work was carried out in part 
under the auspices of the National Nuclear Security Administration of the U.S. Department 
of Energy at Los Alamos National Laboratory under Contract No. DE-AC52-06NA25396.
\end{acknowledgments}

\bigskip % extra skip inserted
% Create the reference section using BibTeX:

\bibliography{MyReferences}

%\begin{thebibliography}{9}   % Use for  1-9  references
%\begin{thebibliography}{99} % Use for 10-99 references
%\bibliography{MyReferences}
%\bibitem{charm07}   http://www.lepp.cornell.edu/charm07/

%\bibitem{templates-ref} http://www.slac.stanford.edu/econf/editors/eprint-template/instructions.html
%\end{thebibliography}

\end{document}